%% file: main.tex
\documentclass[ reprint, superscriptaddress, amsmath,amssymb,
aps, pra, showkeys ]{revtex4-1}

\usepackage[english]{babel}
\usepackage[utf8]{inputenc}
\usepackage{booktabs}   
\usepackage{siunitx}    
\usepackage{array}      

\usepackage[colorinlistoftodos, color=green!40, prependcaption]{todonotes}
\input{preamble}

\usepackage[pdftex, pdftitle={Article}, pdfauthor={Author}]{hyperref} 
\hypersetup{
    colorlinks,
    linkcolor={blue!50!blue},
    citecolor={blue!50!blue},
    urlcolor={blue!80!black}
}

\begin{document}
\title{Temperature-Dependent Emission Spectroscopy of Quantum Emitters in Hexagonal Boron Nitride}

\author{Mouli Hazra}
    \email[Electronic mail: ]{mouli.hazra@tum.de}
\affiliation{Department of Computer Engineering, TUM School of Computation, Information and Technology, Technical University of Munich, 80333 Munich, Germany}
\affiliation{Munich Center for Quantum Science and Technology (MCQST), 80799 Munich, Germany}

\author{Manuel Rieger}
\affiliation{Munich Center for Quantum Science and Technology (MCQST), 80799 Munich, Germany}
\affiliation{Walter Schottky Institute, TUM School of Natural Sciences, Technical University of Munich, 85748 Garching, 
Germany}

\author{Anand Kumar}
\affiliation{Department of Computer Engineering, TUM School of Computation, Information and Technology, Technical University of Munich, 80333 Munich, Germany}
\affiliation{Munich Center for Quantum Science and Technology (MCQST), 80799 Munich, Germany}
   
\author{Mohammad N. Mishuk}
\affiliation{Department of Computer Engineering, TUM School of Computation, Information and Technology, Technical University of Munich, 80333 Munich, Germany}
\affiliation{Munich Center for Quantum Science and Technology (MCQST), 80799 Munich, Germany}

\author{Chanaprom Cholsuk}
\affiliation{Department of Computer Engineering, TUM School of Computation, Information and Technology, Technical University of Munich, 80333 Munich, Germany}
\affiliation{Munich Center for Quantum Science and Technology (MCQST), 80799 Munich, Germany}

\author{Kabilan Sripathy}
\affiliation{Department of Computer Engineering, TUM School of Computation, Information and Technology, Technical University of Munich, 80333 Munich, Germany}
\affiliation{Munich Center for Quantum Science and Technology (MCQST), 80799 Munich, Germany}

\author{Viviana Villafa\~{n}e}
\affiliation{Department of Computer Engineering, TUM School of Computation, Information and Technology, Technical University of Munich, 80333 Munich, Germany}
\affiliation{Munich Center for Quantum Science and Technology (MCQST), 80799 Munich, Germany}
\affiliation{Walter Schottky Institute, TUM School of Natural Sciences, Technical University of Munich, 85748 Garching, 
Germany}

\author{Kai M\"uller}
\affiliation{Department of Computer Engineering, TUM School of Computation, Information and Technology, Technical University of Munich, 80333 Munich, Germany}
\affiliation{Munich Center for Quantum Science and Technology (MCQST), 80799 Munich, Germany}
\affiliation{Walter Schottky Institute, TUM School of Natural Sciences, Technical University of Munich, 85748 Garching, 
Germany}

\author{Jonathan J. Finley}
\affiliation{Munich Center for Quantum Science and Technology (MCQST), 80799 Munich, Germany}
\affiliation{Walter Schottky Institute, TUM School of Natural Sciences, Technical University of Munich, 85748 Garching, 
Germany}

\author{Tobias Vogl}
    \email[Corresponding author: ]{tobias.vogl@tum.de}
\affiliation{Department of Computer Engineering, TUM School of Computation, Information and Technology, Technical University of Munich, 80333 Munich, Germany}
\affiliation{Munich Center for Quantum Science and Technology (MCQST), 80799 Munich, Germany}
\affiliation{Abbe Center of Photonics, Institute of Applied Physics, Friedrich Schiller University Jena, 07745 Jena, Germany}


\begin{abstract}
Color centers in hexagonal boron nitride (hBN) have attracted significant interest due to their potential applications in future optical quantum technologies. For most applications, scalable on-demand fabrication is a key requirement. Recent advances using localized electron irradiation have demonstrated near-identical emitters in the blue and yellow spectral regions. While the blue emitters have been demonstrated in cryogenic temperatures, the yellow emitters remain uncharacterized under such conditions. In this work, we therefore extended the study of yellow emitters to cryogenic temperatures. Initially, multiple spectral features were observed, prompting a systematic investigation that led to the identification of a defect emission centered around 547.5 nm with high brightness and excellent photostability. By tuning the excitation wavelength, we are able to distinguish Raman scattering peaks from the emitter emission. Further analysis of the vibronic emissions allowed us to identify an optical phonon mode, whose contribution becomes increasingly dominant at elevated temperatures. Photoluminescence excitation spectroscopy (PLE) reveals excitation through this phonon mode enhances the emission by almost 5-fold in cryogenic temperature. Temperature-dependent studies further elucidate the role of phonons in the emission process. These observations deepen our understanding of the nature of the emitters, opening new avenues for precise tuning of quantum light sources.
\end{abstract}


\maketitle

\input{sections/Introduction.tex}  

\input{sections/Results_and_Discussions}

\input{sections/Conclusions.tex}

\input{sections/Methods}
\input{sections/acknowledgements.tex}

\input{sections/Author_Declarations}

\bibliography{main}
\appendix*
\input{sections/appendix1.tex}

\end{document}

%% file: preamble.tex
\AtBeginDocument{%
    \newwrite\bibnotes
    \def\bibnotesext{Notes.bib}
    \immediate\openout\bibnotes=\jobname\bibnotesext
    \immediate\write\bibnotes{@CONTROL{REVTEX41Control}}
    \immediate\write\bibnotes{@CONTROL{apsrev41Control,author="48",editor="1",pages="0",title="0",year="1"}}

     \if@filesw
     \immediate\write\@auxout{\string\citation{apsrev41Control}}%
    \fi

}%

\usepackage{caption}
\usepackage[utf8]{inputenc}
\usepackage{amsthm}
\usepackage{mathtools}
\usepackage{physics}
\usepackage{xcolor}
\usepackage{graphicx}
\usepackage[left=23mm,right=13mm,top=35mm,columnsep=15pt]{geometry} 
\usepackage{adjustbox}
\usepackage{placeins}
\usepackage[T1]{fontenc}
\usepackage{lipsum}
\usepackage{csquotes}
\usepackage{comment}

\usepackage{float}
\usepackage{siunitx} 
\usepackage{multirow}
\usepackage{booktabs}
\usepackage{siunitx}
\usepackage{array}
\usepackage{graphicx} 
\usepackage{colortbl}
\usepackage{pifont}
\usepackage{array}

%% file: sections/Introduction.tex
\section{Introduction} \label{sec:introduction}

Color centers in hBN \cite{tran2016quantum} have emerged as an excellent source of single photons for applications in quantum technologies \cite{ahmadi2024quick,ccakan2025quantum,mai2025quantum,cholsuk2024identifying}. Due to the intrinsically large bandgap of approximately 6 eV in hBN \cite{cassabois2016hexagonal}, various color centers can be formed and, in turn, create additional energy levels in the bandgap. Depending on the transition energies between the two-level defect states of each color center, these single photon emitters (SPEs) can potentially generate single photons across a broad spectral range \cite{bourrellier2016bright,shevitski2019blue,cholsuk2022tailoring,cholsuk2024hbn,camphausen2020observation}. Many of these emitters exhibit long-term stability \cite{vogl2018fabrication}, high brightness at room temperature \cite{tran2016quantum}, narrow linewidths \cite{dietrich2018observation}, and weak electron-phonon coupling \cite{li2017nonmagnetic}, making them suitable for integration into quantum photonic devices. In addition, the atomically thin, layered structure of hBN facilitates efficient photon collection \cite{10.1039/C9NR04269E} and integration into on-chip platforms \cite{nonahal2023engineering,li2021integration}. Despite these advantages, preselection of the properties of hBN emitters, such as their emission wavelength, polarization direction, and spatial localization remains challenging. The microscopic nature of many hBN defect centers is to-date poorly understood \cite{cholsuk2023comprehensive}. While the same fabrication techniques can produce multiple species of emitters \cite{hayee2020revealing,tran2016robust}, the emission properties of a single emitter can vary due to external factors such as temperature  \cite{jungwirth2016temperature,korkut2022controlling}, excitation energy \cite{schell2018quantum,shotan2016photoinduced}, or local strain in the crystal lattice \cite{grosso2017tunable}. Therefore, harnessing the full potential of hBN SPEs requires further elucidation of their electronic and optical properties.\\
\indent Electron irradiation using a scanning electron microscope (SEM) has proven to be an effective technique for the site-selective creation of the emitters in hBN. While some studies have demonstrated that electron irradiation can generate blue emitters (called the B center) \cite{fournier2021position,gale2022site,chen2023annealing,zhigulin2023photophysics}, this approach can also be used for the creation of yellow emitters. Notably, blue emitters typically require post-irradiation annealing, whereas yellow emitters can be formed without any pre- or post-treatment \cite{kumar2023localized}. These yellow SPEs are robust and remarkably stable at room temperature (RT). In addition, these emitters exhibit correlated dipole orientations of the excitation and emission axes with respect to the crystal orientation \cite{kumar2024polarization}. While a reproducible emission wavelength at 575 nm has been consistently observed, previous studies \cite{kumar2023localized,kumar2024polarization}  have relied on a fixed excitation wavelength at 530 nm, which resulted in a limited spectral detection range from 550 nm and above, leaving the full emission spectrum only partially explored. In addition, the Raman mode of hBN around 1366 cm$^{-1}$ results in Raman scattering near 575 nm for the 530 nm excitation laser, making the assignment of spectral peaks complicated.\\
\indent In this work, we investigate electron-irradiated hBN at cryogenic temperatures to isolate defect-related emissions from thermal effects and to gain insights into the level structure of the defects. We begin by distinguishing defect-induced emission features from Raman scattering. Focusing on these defect-related transitions, we identify both zero-phonon line (ZPL), corresponding to purely electronic transitions, and vibronic sidebands arising from phonon-assisted processes. A detailed temperature-dependent analysis of these spectral features allows us to quantify the electron-phonon coupling strength, identify the dominant phonon modes involved, and determine the number and nature of phonons participating in the emission process. Our findings provide insights into the optical behavior and possible structural origins of yellow emitters in hBN, advancing their potential use in quantum technology applications.

 \begin{figure*}
    \includegraphics[width = 0.9\textwidth]{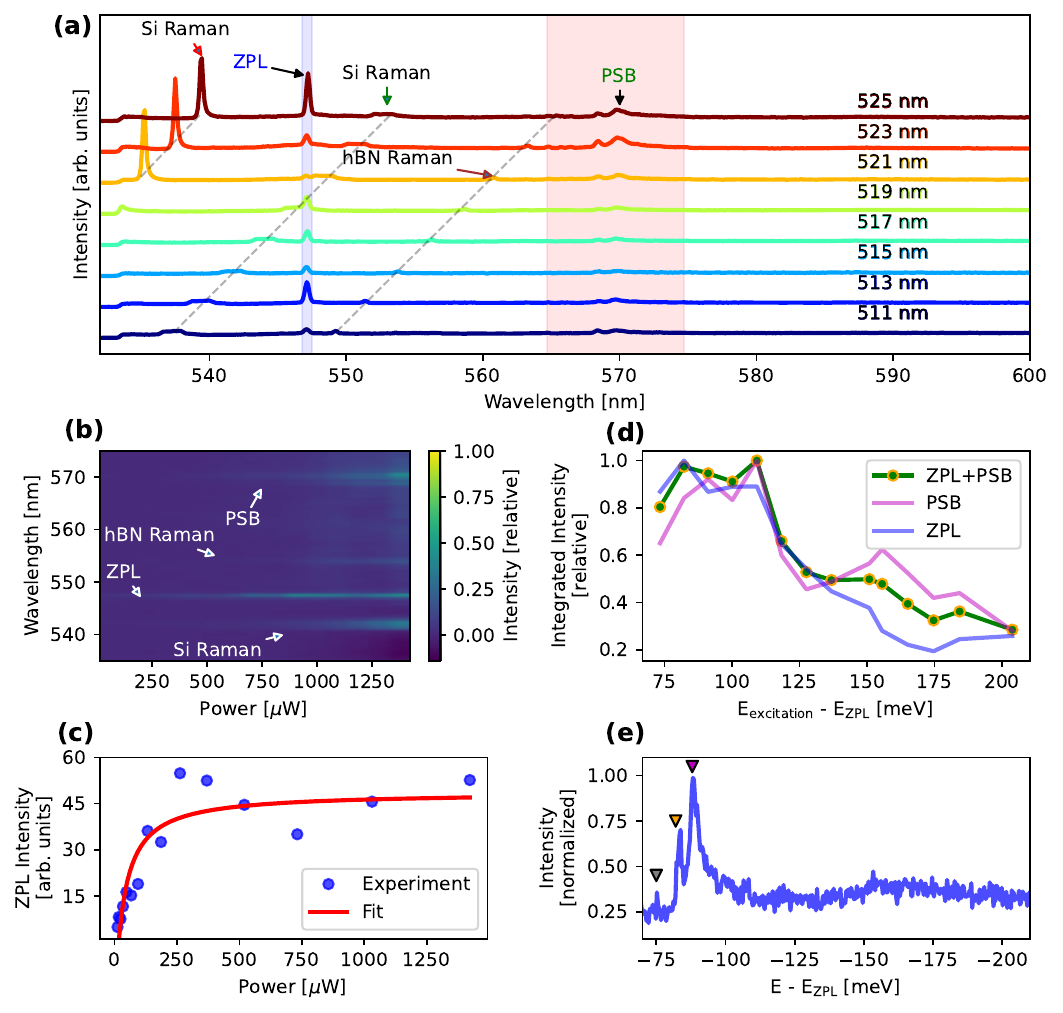}
\captionsetup{
        justification= raggedright, 
        font= small  
    }
    \caption{(a) Low-temperature PL of an hBN emitter excited with a tunable laser. The spectrum features a narrow ZPL at 547.5 nm (blue-shaded region), followed by a broadened PSB at 570 nm (pink-shaded region) and Raman signature from hBN at 1364 cm$^{-1}$ (brown arrow), from the Si substrate at around 517 cm$^{-1}$ (red arrow) and  950 cm$^{-1}$ (green arrow). With varying excitation energy, all Raman peaks shift, while the ZPL and PSB remain at the same positions, changing only in intensity. The spectra were acquired using a tunable continuous-wave (CW) laser at \SI{900}{\micro\watt} excitation power with a 532 nm LP filter in the detection path. (b) A spectrally resolved power dependence PL, showing the ZPL saturates at higher power. (c) The ZPL intensity as a function of excitation power. Fitting into the saturation curve of a two-level system gives saturation power of \SI{34}{\micro\watt}. (d) The PLE spectra, obtained by integrating the PL intensity as a function of excitation energy detuning. The detuning is calculated as the energy difference between the excitation energy and the ZPL energy. The excitation energy at which the emission reaches its maximum closely matches the energy difference between the ZPL and PSB, suggesting a possible phonon-assisted resonant transition. (e) The PSB of a typical emitter is plotted as a function of emission energy detuning with respect to the ZPL energy. The PSB typically peaks around 90 meV but may sometimes split into two, as indicated by the yellow and purple triangles. The grey triangle corresponds to a weaker phonon mode at approximately 74 meV. The PLE spectra were obtained with a tunable 80 MHz pulsed laser with \SI{90}{\micro\watt} excitation power on the sample.}
    \label{Figures/emitter_identification}
\end{figure*}

 

%% file: sections/Results_and_Discussions.tex
\section{Results and Discussions} 
\subsection{Low-temperature Emission in Electron-Irradiated hBN }
Mechanically exfoliated hBN flakes were transferred onto a Si$/$SiO$_2$ substrate and subsequently irradiated with electron beam using a SEM (see Method \ref{sec:Methods}) \cite{kumar2023localized,kumar2024polarization,kumar2024comparative}. The presence of SPEs was initially verified using a commercial photoluminescence (PL) microscope (PicoQuant MicroTime 200) at RT (see Fig.~\ref{PL_Map} in the appendix). The sample was then cooled to 4.5 K in a PL microscope. A tunable continuous-wave (CW) laser was used to excite different regions of the sample. As shown in Fig.~\ref{Figures/emitter_identification}(a), the PL spectrum recorded from a spot above the irradiated region exhibits a sharp, narrow emission at 547.5 nm, accompanied by several additional spectral features. Some of these features, marked with red, green, and brown arrows in Fig.~\ref{Figures/emitter_identification}(a), exhibit an excitation energy-dependent shift, indicating their Raman origin. In contrast, the narrow, intense peak at 547.5 nm, along with a broad emission at 570 nm and smaller features around 566 nm (red-shifted by approximately 90 meV and 74 meV, respectively, from the intense peak), remain unchanged with excitation energy. We assign the 547.5 nm peak to the ZPL and the emission at 570 nm to the phonon sideband (PSB), an attribution that will be justified below.\\
\indent The sharp Raman peak at approximately 517 cm$^{-1}$, along with a broader peak centered around 950 cm$^{-1}$ match well with known Raman modes of Si \cite{borowicz2012deep}. The 517 cm$^{-1}$ Raman peak was blocked by the long-pass (LP) filter for excitations below 519 nm, and we observed this Raman feature across the entire sample. Additionally, a Raman peak at 1364 cm$^{-1}$  (brown arrow) was detected across the entire flake, including at the irradiated spots. This is likely the Raman shift of hBN, which has been reported to be around 1366 cm$^{-1}$ \cite{tran2016quantum}. The slightly lower value observed in our experiments can be attributed to local strain in the sample, which is known to significantly shift the Raman peak in 2D materials \cite{cholsuk2025raman}. Local strain is commonly present in our samples. In our previous work \cite{kumar2024polarization}, we found that both the excitation and emission polarization of the emitters tend to cluster into three groups, as expected for a hexagonal crystal. However, the emission polarization directions were located between the crystal axes rather than aligned with them. These are not separated by the expected 60°, indicating the presence of local strain that can break the crystal symmetry. Notably, at the 530 nm excitation wavelength, the Raman from hBN coincides with the 570 nm peak that was reported in our previous works \cite{kumar2023localized,kumar2024polarization}. By choosing the excitation wavelength, we can therefore prevent any coincidence of Raman scattering with the emission from the irradiated spots for further investigations.\\
\indent In the next step, we excited the same spot using a 515 nm diode laser and recorded the emission as a function of excitation power. The results as presented in Fig.~\ref{Figures/emitter_identification}(b) revealed that the 547.5 nm emission is the most dominant at low excitation power. As the power increases, additional emission peaks appear, with all features initially following a linear power dependence. Fig.~\ref{Figures/emitter_identification}(c) shows the integrated peak intensity of the 547.5 nm emission peak is plotted after background subtraction for each excitation power. It saturates at higher excitation power, confirming a defect-related origin \cite{goppert1931elementarakte}. Other spectral features marked with red, green, and brown arrows in Fig.~\ref{Figures/emitter_identification}(a), exhibit a continuous linear increase with power (see Fig.~\ref{Power_Series}), further confirming their assignment as Raman modes. The saturation behavior closely fits with the two-level saturation model $I_{\text{ZPL}}(P) = I_{\text{sat}} \frac{P}{P + P_{\text{sat}}} + I_{\text{d}}$ with saturation intensity, $I_{\text{sat}} = $ 34 $\mu$W. Notably, the ZPL intensity appears to oscillate around the saturation level at higher excitation powers, which may be indicative of complex emission dynamics. This should be investigated further in future work. \\
\indent We next integrate over the assigned ZPL and PSB peaks as a function of excitation energy. The shaded regions in Fig.~\ref{Figures/emitter_identification}(a) mark the spectral integration bandwidth. If the peak at 547.5 nm is ZPL and 570 nm PSB (energy difference of 90 meV), then the excitation should be very efficient at ZPL plus 90 meV (which corresponds to 526 nm), when the excess energy of the laser corresponds exactly to the phonon mode. As shown in Fig.~\ref{Figures/emitter_identification}(d), the (normalized) intensity of the peaks indeed peaks at around 90 $\pm$ 20 meV energy difference and is significantly enhanced by a factor of five. For clarity, we have also shifted the PSB relative to the ZPL in Fig.~\ref{Figures/emitter_identification}(e), which then peaks at $\sim$90 meV. The relatively broader peak in the PLE spectra around 90 meV likely arises from the finite linewidth of the excitation laser. We therefore can conclude that our assignment of ZPL and PSB is justified.\\
\indent The defect-related emissions were detected only on and around the irradiated spots, suggesting that these defects are created exclusively at locations where the electron beam interacts with the hBN flake. Overall, excitation energy-dependent and power-dependent PL measurements allow for the systematic identification of defect-related emission while enabling clear differentiation from Raman signals originating from the substrate and host hBN crystal.
 
\subsection{Photophysical Properties at Cryogenic Temperature}
\begin{figure*}
    \includegraphics[width = 1\textwidth]{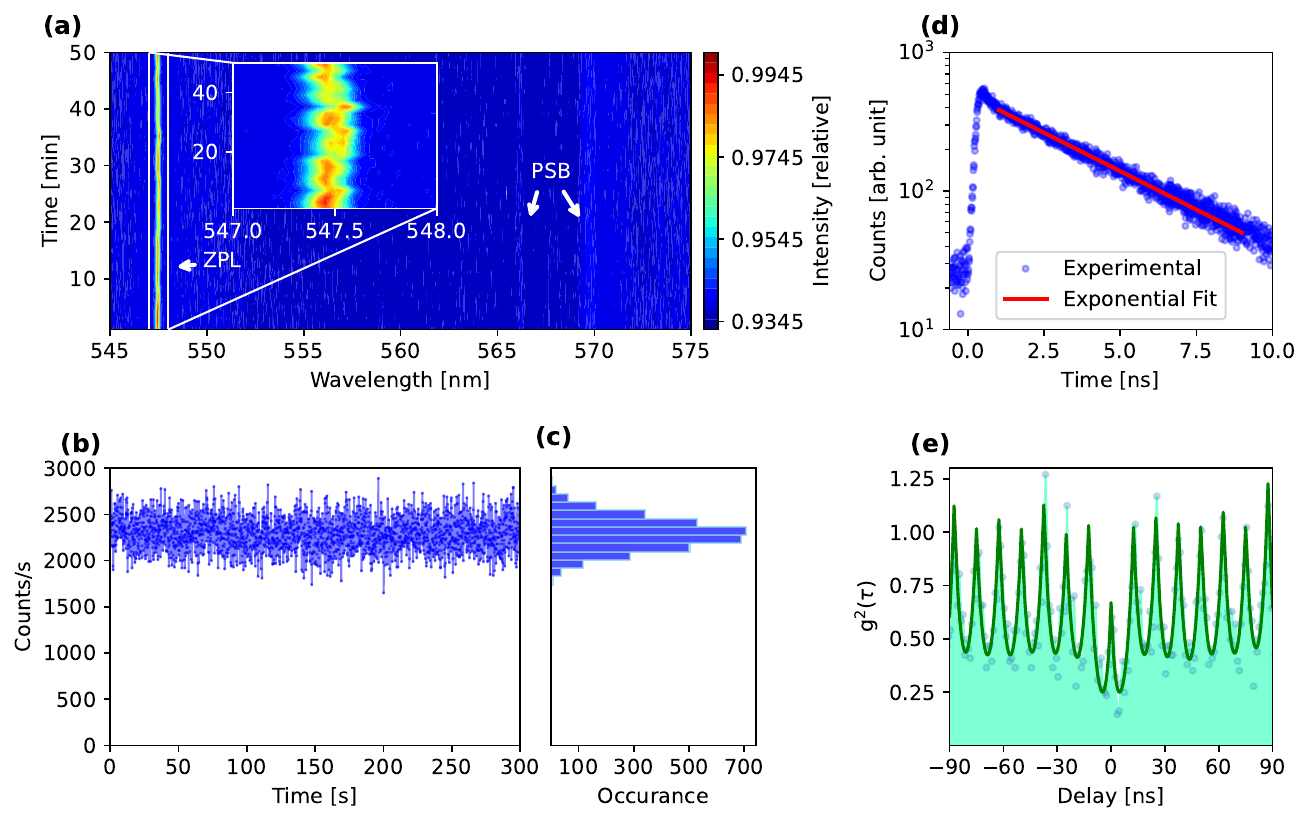}
\captionsetup{
        justification= raggedright, 
        font= small  
    }
    \caption{Optical properties of hBN emitters at 4.5 K. (a) Time-resolved emission spectra reveal spectral diffusion of $ \pm$ 0.2 nm, with the mean ZPL centered at 547.5 nm as shown in the inset. Each spectrum was acquired with a 10 s integration time and averaged over six measurements with \SI{190}{\micro\watt} power on the sample. The spectra are normalized to the maximum intensity. (b,c) The Photon count under continuous excitation depicts the stability of the emitter, showing no signs of blinking or bleaching throughout the 5-minute observation period. (d) Excited state lifetime measurement yields a lifetime of $\tau =$  3.915$ \pm$ 0.024 ns. (e) Second-order correlation measurement $g^{2}(\tau)$ revealing a dip (0.505) at zero delay time. The fitting is only for the guide to the eye. All the measurements were done with a 515 nm pulsed laser with a 80 MHz repetition rate.  }
    \label{Figures/photophysical_prop}
\end{figure*}

\indent After identifying the defect-related emission, we investigated its photophysical properties at low temperature. The emitter was continuously excited non-resonantly with an 80 MHz pulsed laser. The resulting emission spectra  over a 50-minute period are presented in Fig.~\ref{Figures/photophysical_prop}(a). The ZPL exhibited spectral diffusion around 547.5 $\pm$ 0.2 nm as depicted in the inset. Integrating the spectra for all time steps and fitting it in a Gaussian yields a linewidth of 260 GHz. Importantly we note that no lines appear/ disappear, which would reveal additional energy levels or more complicated photodynamics. Meanwhile, time-resolved photon counting over 5 minutes revealed no signs of blinking or bleaching, as can be seen in Fig.~\ref{Figures/photophysical_prop}(b,c). The mean count rate is 2290 $\pm$ 172 counts/s, which remains above the shot noise limit (the square-root of the mean is 48 counts/s).\\
\indent To further characterize the emitter, we measured its excited-state lifetime, as shown in Fig.~\ref{Figures/photophysical_prop}(d). A single-exponential fit to the decay curve yielded a lifetime of 3.92 $\pm$ 0.02 ns, which is consistent with previous reports for hBN emitters \cite{korkut2022controlling,guo2023coherent,aharonovich2016solid} and our previous measurements \cite{kumar2023localized}. A  similar lifetime (3.68$\pm$0.02 ns) was observed when using a 570 (10) nm bandpass (BP) filter, which would be expected for a PSB. \\
\indent To assess the quantum nature of the emission, second-order intensity correlation measurements ($g^{(2)}(\tau)$) were performed. The results reveal a distinct dip at zero time delay as depicted in Fig.~\ref{Figures/photophysical_prop}(e). We calculated the ratio of the peak area at zero time delay to the average peak area, which resulted in a value of $g^{(2)}(0) =$ 0.505. A value of $g^{(2)}(0) <$ 0.5 is typically considered indicative of a single photon emitter \cite{grunwald2019effective}, as the overlap with the single photon Fock state is non-zero. However, the slightly higher value observed here can be understood in the context of the experimental conditions. The setup did not allow for spatial PL mapping, which limited our ability to find an isolated emitter and accurately align the excitation laser with it. As a result, the collected signal may include contributions from two emitters within the diffraction-limited spot. For example, if there are two emitters ($N_{\text{emitter}} = 2$), then $g^{(2)}(0) = 1 - 1/N_{\text{emitter}} = 0.5$. Additionally, the Raman peaks present in the spectra are all coupled to the single photon detectors. Since these peaks inherit Poissonian statistics from the excitation laser, they contribute to an increased $g^{(2)}(0)$. Overall, the emitter exhibits a ZPL with well-separated PSB, demonstrating stable emission without noticeable blinking or significant spectral diffusion at cryogenic temperatures. 

\subsection{Temperature-Dependent Photoluminescence of the Emitter}
 \begin{figure*}
 \includegraphics[width = 1\textwidth]{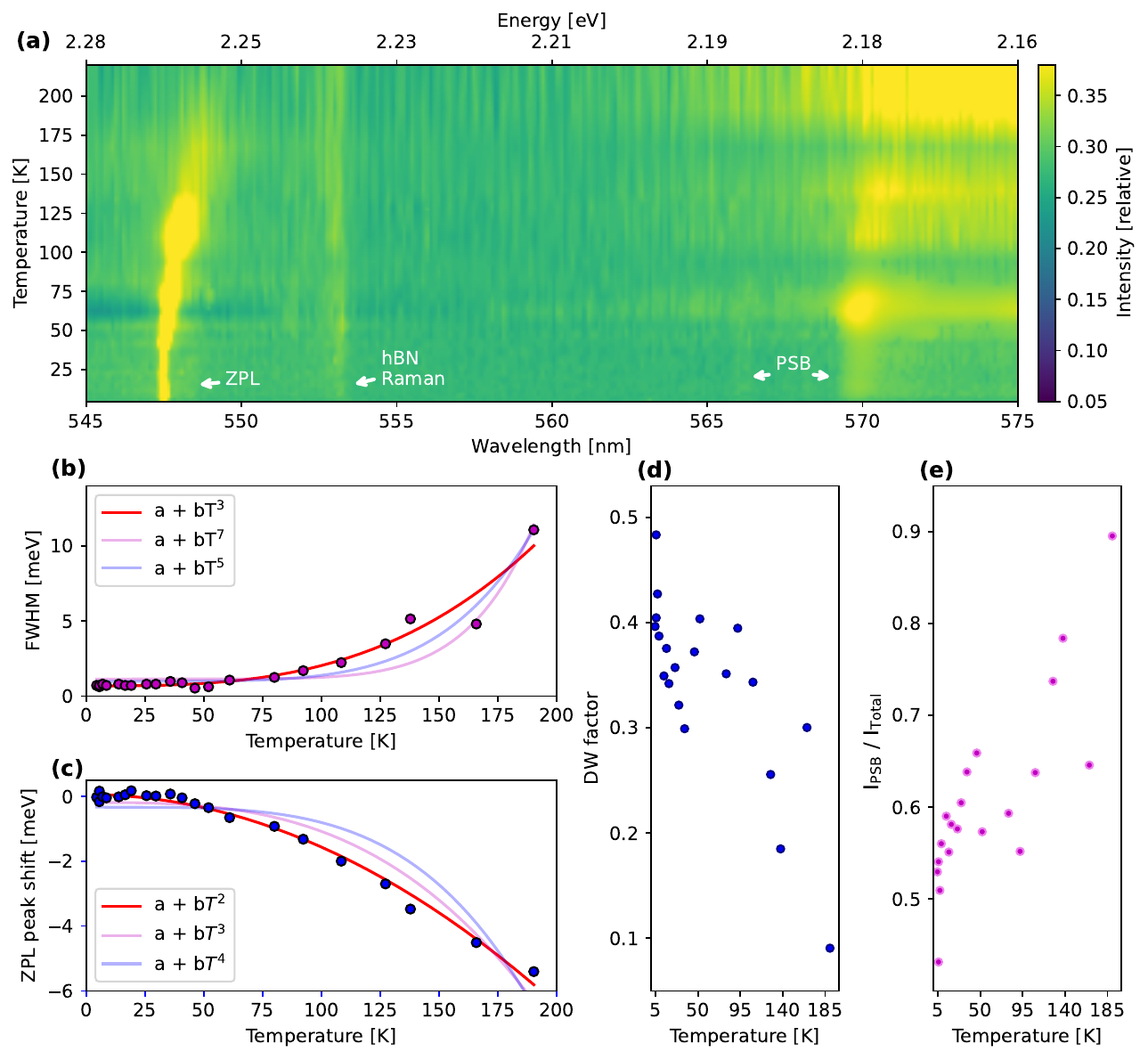}
 \captionsetup{
        justification= raggedright, 
        font= small  
        }
        \caption{(a) Temperature-dependent photoluminescence excited with a 515 nm diode laser. (b) Energy shift and (c) linewidth of ZPL centered around 547.5 nm (2.2645 eV) with a linewidth of 7.8 meV at 4.5K, extracted after fitting the ZPL with a Lorentzian. The phonon contribution gets dominant with temperature as indicated by the (d) Debye-Waller (DW)  factor and (e) ratio of the integrated PSB and total intensity. The emission spectra were recorded by exciting the emitter with a 515 nm diode laser with \SI{970}{\micro\watt} power on the sample and using a fine spectrometer grating with 300 lines/mm for a 10 second integration time, averaging 6 spectra. The background was subtracted from each temperature step before normalizing them.}
        \label{Figures/temp_series}
\end{figure*}
\indent After investigating the cryogenic properties of the emitters, we further investigate it's emission properties and how they evolve with temperature to gain insights into the nature of the electron-phonon coupling. We excite the emitter with a 515 nm diode laser and systematically vary the temperature. Fig.~\ref{Figures/temp_series}(a) presents the PL spectra recorded over the temperature range from 4.5 K to 220 K. At each temperature, the background signal was subtracted and the spectra were normalized to the maximum intensity before being plotted in a contour plot. At low temperatures, the PL spectrum exhibits a sharp ZPL centered at 547.5 nm (2.26 eV) with a linewidth of 7.8 meV at 4.5 K, accompanied by a prominent PSB at approximately 90 meV and a weaker PSB 74 meV redshifted from the ZPL. As the temperature increases, the ZPL exhibits both spectral broadening and a redshift, while the PSB increases in intensity and becomes broader. The separation between the ZPL and PSB reduces with temperature, but the PSB becomes stronger on the longer-wavelength side compared to the shorter-wavelength side. Furthermore, PSB features at $\sim$74 meV and $\sim$90 meV ($\sim$566 nm and 570 nm in Fig.~\ref{Figures/temp_series}(a)) gradually merge at elevated temperatures, indicating temperature-dependent changes in electron–phonon coupling. Above 130 K, the relative intensity of the ZPL compared to the background decreases substantially, and the ZPL becomes undetectable beyond 200 K. Notably, the Raman signal from bulk hBN increases slightly in intensity and broadens with temperature, and the background signal remains nearly constant across the temperature range, confirming that they are unrelated to the quantum emitter. \\
\indent To quantitatively analyze the temperature-induced changes in the ZPL, the emission peak was fitted using a Lorentzian function (see Fig.~\ref{emission_stat}), and both the full-width at half-maximum (FWHM) and the central wavelength were extracted. The results of the fitting are presented in Fig.~\ref{Figures/temp_series}(b) and (c) respectively. In the low-temperature regime (up to approximately 50~K), the ZPL linewidth remains nearly constant, suggesting that dephasing is dominated by temperature-independent mechanisms. Beyond that, the ZPL linewidth increases with temperature, as depicted in Fig.~\ref{Figures/temp_series}(b). Similarly, the ZPL emission energy remains nearly unchanged below 50 K and begins to redshift by approximately 6 meV between 4.5 and 200 K. This shift can be attributed to reduced bandgap at elevated temperature \cite{o1991temperature,du2014temperature,tan2022donor}. To understand the underlying phonon contributions, the temperature dependence of the FWHM and peak energy of ZPL was fitted to various power-law functions. In the intermediate temperature range (approximately 50--150~K), the FWHM follows a $T^3$ trend that has been observed previously in hBN \cite{jungwirth2016temperature,tan2022donor,korkut2022controlling,white2021phonon,sontheimer2017photodynamics} and other color centers \cite{neu2013low,mueller2012phonon,lienhard2016bright,hizhnyakov2002zero}. Meanwhile, the ZPL redshift above 50~K is well described by a $T^2$ dependence. These temperature dependencies deviate from the predictions of standard harmonic models \cite{maradudin1966theoretical}. Specifically, quadratic electron-phonon coupling under the Debye approximation anticipates a $T^7$ dependence for linewidth broadening and a $T^4$ dependence for ZPL redshift. Furthermore, an increase in linewidth at higher temperatures slower than $T^5$ indicates that dynamic Jahn-Teller effects are unlikely to play a significant role \cite{fu2009observation}. Taken together, these deviations suggest alternative dephasing mechanisms. Notably, the observed $T^3$ linewidth broadening and $T^2$ redshift are consistent with soft-mode theory \cite{hizhnyakov2002zero,mueller2012phonon,razgulov2021low,neu2013low}. In this framework, electronic excitation weakens the local bonding, increasing the density of low-frequency phonon modes \cite{hizhnyakov2002zero}. As temperature rises, the Bose–Einstein occupation of these phonon modes increases, enhancing their coupling to the electronic states. This leads to temperature-dependent dephasing and energy renormalization, observed as ZPL broadening and redshift.\\
\indent As the ZPL broadens and shifts to longer wavelengths, the PSB undergoes changes in both shape and intensity with increasing temperature. To quantify electron-phonon coupling through the PSB pathway, we calculated the Debye-Waller (DW) factor, defined as the ratio of the integrated ZPL intensity to the total integrated PL intensity, as presented in Fig.~\ref{Figures/temp_series}(d). Below 50 K, the DW factor remains relatively high ($\sim$0.5), indicating weaker electron-phonon interactions. As temperature increases, the DW factor decreases, reflecting the growing contribution of the phonon-assisted emission process. This trend can be explained by the fact that at low temperatures, emission predominantly occurs through the ZPL pathway, whereas at higher temperatures, thermally excited phonons increase PSB contributions. This leads to a rise in the ratio of the integrated PSB intensity to the total PL intensity ($I_{\text{PSB}}/I_{\text{tot}}$) within the 545 nm to 575 nm range as the temperature increases, as depicted in Fig.~\ref{Figures/temp_series}(e). Above 200 K, the ZPL is no longer observable, while the PSB intensity continues to increase with temperature. 

\subsection{Vibronic Emission}
During the light emission process, the transition from an excited electronic state to a lower-energy ground state may be purely radiative or vibronic when phonons are involved. Building upon the experimental observations and discussions presented above, we aim to gain further insight into the phonon modes involved in these vibronic transitions, the strength of their coupling, and their possible origins.\\ 
\indent In the PLE spectra [Fig.~\ref{Figures/emitter_identification}(d)], we identified a phonon mode with an energy of approximately 90 meV, which appears to participate in both absorption and emission processes. This mode consistently dominates the PSB and exhibits slight variations in spectral shape across different emitter sites, reflecting sensitivity to the local environment. One-phonon spectra extracted at 4.5 K (see Fig.~\ref{PSB_temp}) reinforce the dominance of this 90 meV mode, suggesting it is the primary vibronic contributor within the observed energy range. In some sites, two distinct peaks emerge within this region, as shown, for example, by Fig.~\ref{Figures/emitter_identification}(e) and Fig.~\ref{PL_Map}(c). As depicted by the power-dependent measurements in Fig.~\ref{Figures/emitter_identification}(b), the lower-energy peak becomes visible only at higher excitation powers, indicating that the coupling efficiency may vary depending on the emitter site. A weaker emission feature near 74 meV is occasionally observed in the vibronic spectra. At low temperatures, it remains close to the noise level, and at higher temperatures, it merges with the dominant PSB. Due to its relatively lower energy difference and proximity to the dominant optical mode, it may represent a weakly coupled component of the same optical phonon mode. \\
\indent In addition, the ZPL linewidth exhibits a $T^3$ temperature dependence, while its energy undergoes a redshift that follows a $T^2$ trend. These behaviors are consistent with soft-mode dynamics, in which low-energy phonons become increasingly active at elevated temperatures \cite{hizhnyakov2002zero}. This interpretation appears to contrast with the PSB structure, which is dominated by a high-energy ($\sim$90 meV) phonon mode at elevated temperature. This discrepancy suggests that the phonons responsible for ZPL broadening and energy renormalization are distinct from those contributing to the vibronic sideband. It implies the presence of additional low-energy phonon modes that influence ZPL dephasing but remain weakly coupled to the optical transition and are therefore not prominent in the emission spectrum. Overall, our investigation into  the  role of phonon interactions in the emission lineshapes provides valuable insight into the vibronic structure of the hBN quantum emitter created by electron beam irradiation.

%% file: sections/Conclusions.tex
\section{Conclusions}\label{sec:conclusions}
\indent In summary, we investigated low-temperature PL from exfoliated hBN samples irradiated with electron beam using an SEM. By systematically varying the excitation power and photon energy of the laser, we identified defect-related emission at cryogenic temperature. Furthermore, the use of tunable excitation allowed us to distinguish between Raman scattering and defect-related luminescence, enabling more accurate attribution of spectral features. Following the identification of the defect-related emission, we characterized the photophysical properties of the emitters. The emission remained stable over time, showing minimal blinking and spectral diffusion under cryogenic conditions. Additionally, we investigated the temperature-dependent  PL of electron-irradiated hBN to gain insights into the electron-phonon interactions governing the emission properties. Similar studies have previously been conducted on hBN quantum emitters, revealing various temperature-dependent behaviors of ZPL linewidth, such as $T^3$ \cite{white2021phonon,sontheimer2017photodynamics,tan2022donor}, $ T + T^5$ \cite{ari2025temperature}, nearly exponential \cite{jungwirth2016temperature}, or linear trends \cite{akbari2021temperature} implying different underlying physical mechanisms. The emitters in those studies were typically created/activated via annealing drop-casted hBN nanoflakes in ethanol/water solution in N$_2$ environment \cite{sontheimer2017photodynamics,tan2022donor,jungwirth2016temperature}, by exposing exfoliated hBN to H$_2$ \cite{white2021phonon} plasma followed by annealing in air and annealing in an Ar environment \cite{akbari2021temperature}. Differences observed across the literature are likely attributable to the wide range of structural and electronic configurations of defect centers, highlighting the intrinsic heterogeneity of color centers in hBN. While low-temperature studies on hBN emitters created via electron beam irradiation have been reported \cite{gale2022site,fournier2021position,zhigulin2023stark}, our result represents a comprehensive temperature-dependent investigation of their optical properties. Deeper insight into these systems is essential for the development of quantum technologies, where characteristics such as emission linewidth, spectral stability, and photon indistinguishability are important. In addition, identification of the phonon mode, together with PLE measurements, revealed an efficient excitation mechanism via optical phonon-assisted absorption, leading to a nearly fivefold enhancement in emission. Despite the presence of strong electron-phonon coupling, which tends to suppress the ZPL at elevated temperatures, the emitters demonstrated high single-photon purity even at room temperature, as indicated by low $g^{2}(0)$ values in our previous work \cite{kumar2023localized,kumar2024polarization}. This makes them promising candidates for quantum key distribution systems \cite{abasifard2024ideal}, where both emission probability and photon purity are critical.\\
\indent Beyond technological relevance, accurate identification of the ZPL and PSBs also provides valuable clues for determining the microscopic origin of the defects. Such assignments are often supported by comparisons with density functional theory (DFT) calculations \cite{cholsuk2023comprehensive,kumar2023localized}. However, due to inherent limitations in the accuracy of DFT, particularly its typical assumption of 0 K temperature conditions \cite{cholsuk2023comprehensive,cholsuk2024hbn}, it is crucial to combine theoretical insights with low-temperature experimental data to obtain a fuller understanding. Our findings provide a valuable experimental foundation for guiding and refining future theoretical models, improving the identification and classification of defects in electron-irradiated hBN, and opening new opportunities for defect engineering in emerging quantum technology applications.

%% file: sections/Methods.tex
\section*{Methods} \label{sec:Methods}

\subsection{Emitter Creation}
\indent Emitters were created by electron irradiation in an exfoliated hBN flake with a thickness of approximately 30–60 nm, using a Helios NanoLab G3 system, as described in previous work \cite{kumar2023localized}. The irradiation was performed with 7.7 $\cross$ 10$^{17}$ cm$^{-2}$ fluence, 3 kV acceleration voltage, 25 pA current, and 10 seconds dwell time. No post-treatment was required.
\subsection{Optical Characterization}
\indent The sample was placed inside a cryostat (attoDRY 800, Attocube Inc.), mounted on a piezo positioner (AMC300 piezo driver with ANPx311 nanopositioner for three different axes). Excitation was provided by a fiber-coupled tunable CW laser (C-Wave GTR), which was passed through two shortpass (SP) 550 nm filters and a 515 (30) BP filter  to ensure spectral purity. A 0.8 NA apochromatic objective was used to focus the excitation laser onto the sample as well as to collect the emission. The emission was subsequently filtered using a 532 nm LP and 600 nm SP filter, with additional BP filters applied where necessary. The filtered emission was coupled into a fiber and recorded using a Shamrock 500i spectrometer equipped with an iDUS-416 CCD camera. A non-polarizing beamsplitter (BS025, Thorlabs) was used to measure the transmitted laser power before entering the cryostat. The actual power incident on the sample was determined using the reflectance datasheet for the beamsplitter. For power-dependent measurements, an electric variable optical attenuator (V450PA, Thorlabs) was introduced at the end of the excitation fiber. The PLE measurements were performed using both pulsed and CW lasers, with the CW laser operated at 900 µW, exceeding the saturation power. However, due to the broad spectral width of the pulsed laser, detecting the Raman signal was challenging.
\indent For temperature-dependent characterization, the excitation source was a 515 nm single-mode diode laser (iBeam). A 70:30 beamsplitter (BS022, Thorlabs) was used to direct the excitation beam onto the sample and measure the power.
\subsection{Time-Resolved Measurements}
\indent For time-resolved PL measurements, the emitters were excited using an 80 MHz pulsed laser with 100 fs pulse width, and the collected signal was directed to a superconducting waveguide single-photon detector (Single Quantum Inc.). For lifetime measurements, the detector was triggered by the pulsed laser, which initiated histogram recording. For second-order correlation measurements, the emission was split using a fiber-coupled 50:50 beamsplitter (PN560R5A1, Thorlabs) before being sent to the detectors. The pulsed laser undergoes temporal broadening due to fiber dispersion that was observed in the second order autocorrelation measurement.

%% file: sections/acknowledgements.tex
\section*{Acknowledgements} \label{sec:acknowledgements}
 This research is part of the Munich Quantum Valley, which is supported by the Bavarian state government with funds from the Hightech Agenda Bayern Plus. This work was funded by the Deutsche Forschungsgemeinschaft (DFG, German Research Foundation) under Germany's Excellence Strategy - EXC-2111-390814868 (MCQST) and as part of the CRC 1375 NOA project C2 and project  PQET (INST 95/1654-1). The authors acknowledge support from the Federal Ministry of Research, Technology and Space (BMFTR) under grant number 13N16292 (ATOMIQS) and by the German Space Agency DLR with funds provided by the Federal Ministry for Economic Affairs and Climate Action BMWK under grant numbers 50WM2165 (QUICK3), and 50RP2200 (QuVeKS).

%% file: sections/Author_Declarations.tex
\section*{Notes}
We acknowledge the use of AI assistance for support in data analysis, visualization, and rephrasing parts of the manuscript for clarity. All scientific interpretations, data analyses, and final writing decisions were made by the authors. The authors declare no competing  financial interest. 

 

%% file: sections/appendix1.tex
\clearpage
\renewcommand{\thefigure}{A\arabic{figure}}
\setcounter{figure}{0}
\begin{appendix}

\onecolumngrid

\section{Photoluminescence Map of electron irradiated hBN}
\begin{figure}[H]
    \centering
    \includegraphics[width=0.9\textwidth]{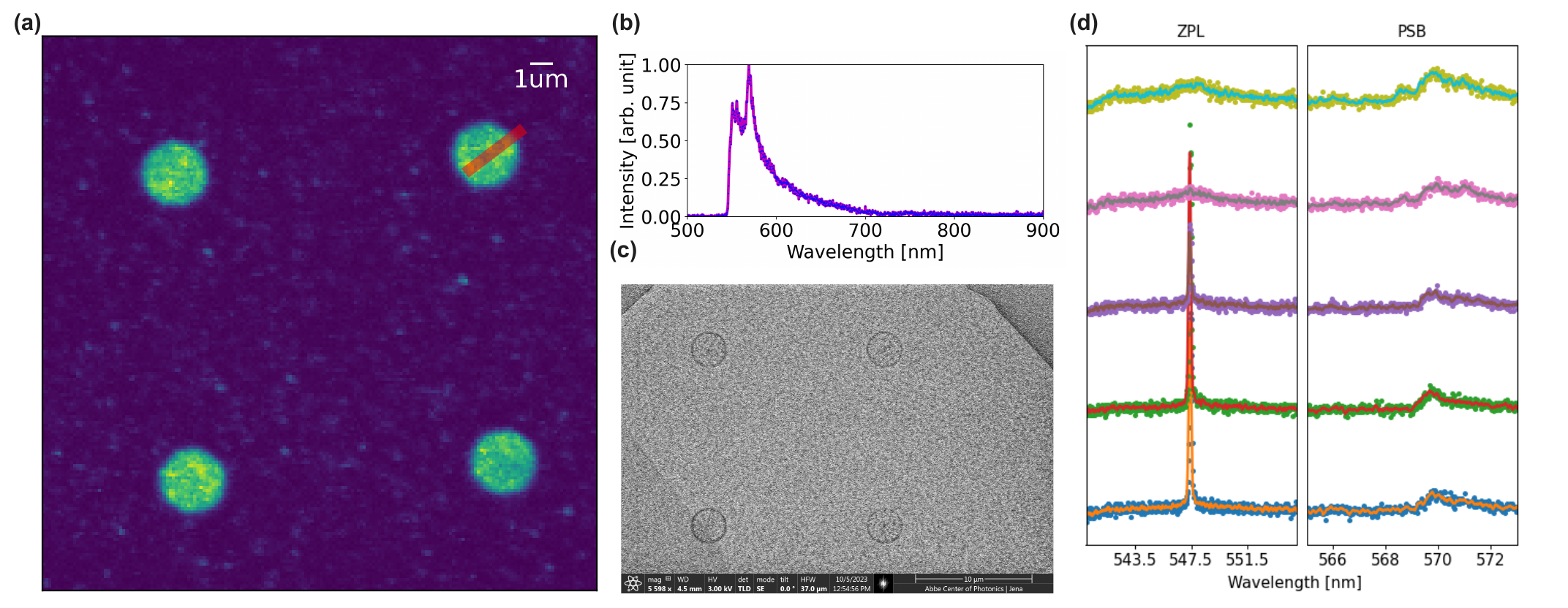}
    \captionsetup{justification=raggedright, font=small}
    \caption{(a) Room temperature PL map from electron-irradiated hBN using a 532 nm pulsed laser with a 550 nm long-pass (LP) filter in the emission path.
    (b) PL spectra from an emitter on the irradiated spot at room temperature. (c) SEM image of the same flake after electron irradiation.(d) Low-temperature (4.5 K) PL spectrum recorded from the same flake along the red line shown in Fig. (a), with 515 nm pulsed laser.}
     \label{PL_Map}
\end{figure}

\section{Power Dependence of Raman Signals}
\begin{figure}[H]
    \centering
        \includegraphics[width = 0.5\textwidth]{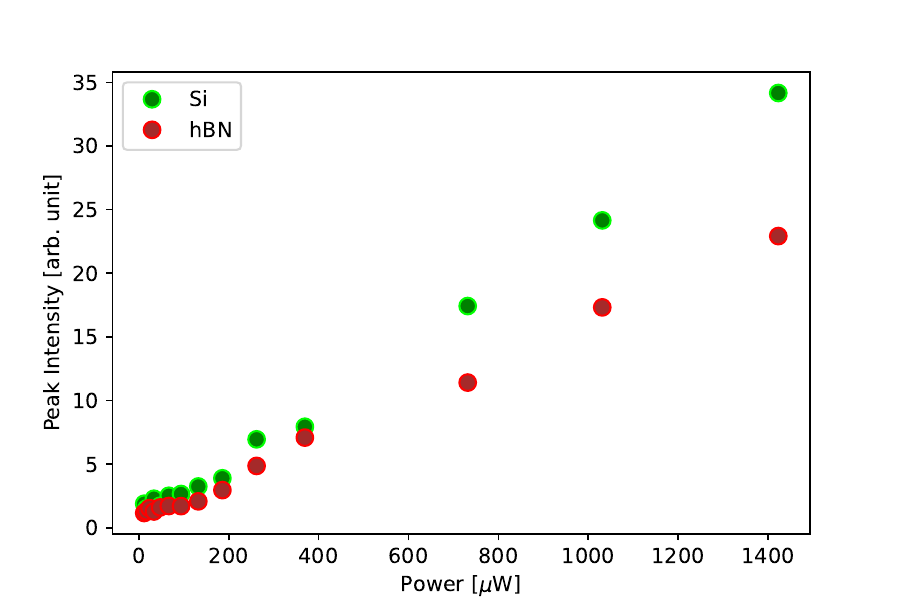}
    \captionsetup{
        justification=raggedright, 
        font= small  
    }
    \caption{Peak intensity of the Raman peak of Si at 950 cm$^{-1}$ and hBN at 1364 $^{-1}$ (corresponding to the peak around 542 nm and 553 nm in Fig.~\ref{Figures/emitter_identification}(b)) as a function of excitation power after background sunstraction. Here the background is defined as the mean spectral count between 550 nm and 560 nm, as this region remains the flattest across all measurements.}
    \label{Power_Series}
\end{figure} 

\section{Emission Statistics}

     \begin{figure}[H]
    \centering
        \includegraphics[width = 0.9\textwidth]{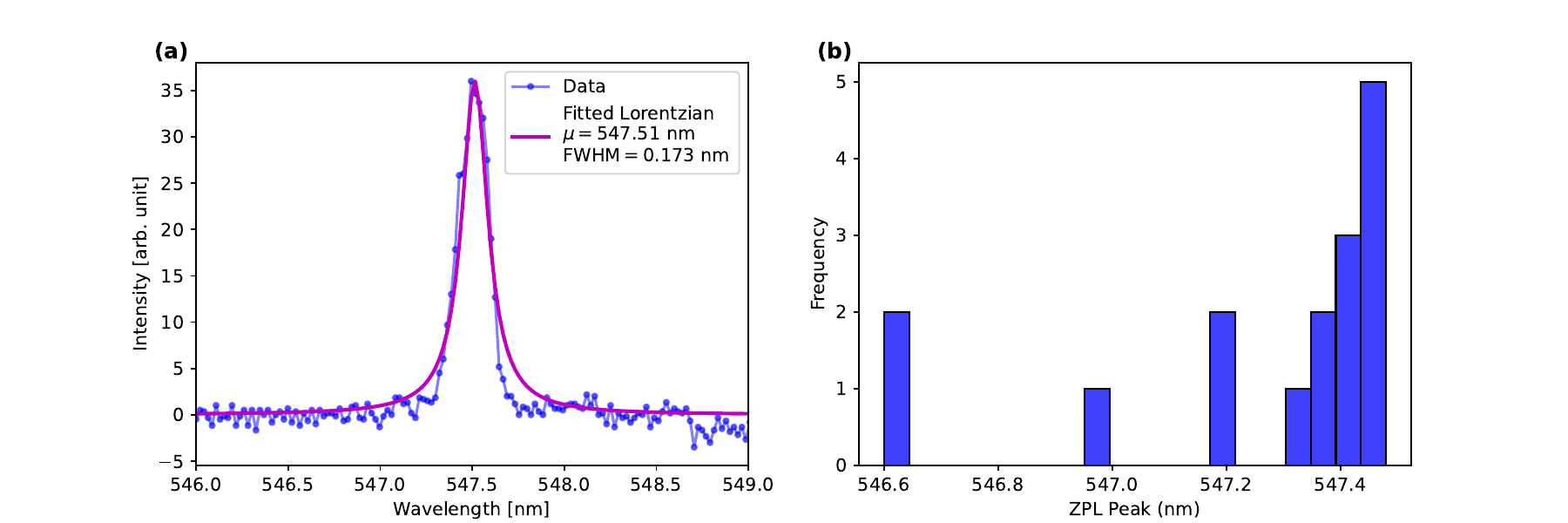 }
        \captionsetup{width=1\textwidth, justification=raggedright, font=small}
    \captionsetup{
        justification=raggedright, 
        font= small  
    }
	\caption{(a) The ZPL was fitted with a Lorentzian : $
f(x) = \frac{a}{1 + \left(\frac{x - \mu}{\gamma}\right)^2} $$
$ with FWHM defined as $2\gamma $, where $\gamma$ represents the standard deviation of the Lorentzian fit. Notably at low temperatures, subtle variations of FWHM on timescales shorter than the spectrometer integration time and/or limited by spectrometer resolution remain undetected.(b) Distribution of ZPL emission wavelength from 16 different spots in and around the irradiated regions. The majority of emissions are centered around 547.5 nm, with slight variations likely attributed to spectral diffusion, as discussed in Fig. 2 of the main text. Additionally, two emitters exhibit ZPLs at 546 nm, This could be related to strain, but further investigation is still required.}
\label{emission_stat}
\end{figure}


\section{PSB Analysis}
In order to assess the impact of lattice vibrations on the emission spectrum, we follow the standard Huang–Rhys framework \cite{maradudin1966theoretical,davies1974vibronic,exarhos2017optical}, which decomposes the total lineshape $L(E)$ into a zero‐phonon component and its vibronic sideband.  In energy units, this can be expressed as
\begin{equation}
    L(E) = e^{-S_\text{HR}}I_0(E) + I_0(E) \otimes I_\text{PSB}(E),
\end{equation}
where $S_\text{HR}$ is the Huang-Rhys factor. $I_0(E)$ denotes the normalized Lorentzian profile fitted to the ZPL from an experiment. $I_\text{PSB}(E)$ is the PSB lineshape and can be constructed as a Poisson-weighted sum over $n$-phonon processes, as shown below
\begin{equation}
    I_\text{PSB}(E) = \sum_n e^{-S_\text{HR}}\frac{S_\text{HR}^{n}}{n!}I_n(E).
\end{equation}
Here, $I_n(E)$ is the normalized $n$-phonon probability distributions where $I_n(E) = I_1 \otimes I_{n-1}$ where $n \geq 2$. This implies that each $n$-phonon distribution arises by successive convolution of the single-phonon distribution ($I_1(E)$). In this work, up to 20 phonons were considered. At finite temperatures, the one‑phonon term is governed by Bose-Einstein statistics; therefore, we assume the following function, similar to Ref.~\cite{exarhos2017optical}, to describe the single-phonon distribution
\begin{equation}
    I_{1}(E) = 
                    \begin{cases} 
                        A\left[\frac{1}{e^{E/k_{B}T} - 1}+1\right]S(E) & \text{for }  E>0 \\ 
                        \\
                        A\left[\frac{1}{e^{-E/k_{B}T} - 1}\right]S(-E) & \text{for }  E<0 .
                    \end{cases}
\end{equation}
Here, $A$ is a normalization term. $k_B$ is the Boltzmann constant. $T$ is a temperature, which we use the same value as in our experiment. The function \(S(E)\) is the phonon spectral function, which we represent by discrete values on a uniform grid from 0 to 200 meV with 0.5 meV spacing. These sets of \(\{S(E_{i})\}\) are the fitting parameters. We then adjust \(\{S(E_{i})\}\) along with \(S_{\rm HR}\), \(E_{\rm ZPL}\), and the FWHM of the fitted Lorentzian lineshape from ZPL (\(\Gamma_{\rm ZPL}\)) until \(L(E)\) matches an experimental lineshape (\(L(E)_{\rm exp}\)) over the entire energy grid. This \(L(E)_{\rm exp}\) can be acquired from
\begin{equation}
L(E)_{exp} = \frac{S(\lambda)hc}{E^5}.
\end{equation}
Here, $S(\lambda)$ is a PL intensity measured from our experiment as a function of wavelength. We then convert it to the energy unit by applying the Jacobian transformation \(\lvert d\lambda/dE\rvert=hc/E^{2}\) and accounting for the spontaneous emission by the scaling of $E^3$ \cite{exarhos2017optical}.

As revealed by Ref.~\cite{exarhos2017optical}, the Huang-Rhys formalism tends to exhibit an artifact in the low-energy phonon regime. Therefore, we impose an artificial linear constraint on $S(E)$ until $E < 5$ meV to prevent this contribution to the Huang-Rhys factor. The upper limit of 200 meV coincides with the maxima of the phonon density of states in bulk hBN \cite{tohei2006debye}.

\begin{figure*}
    \centering
    \includegraphics[width=0.5\linewidth]{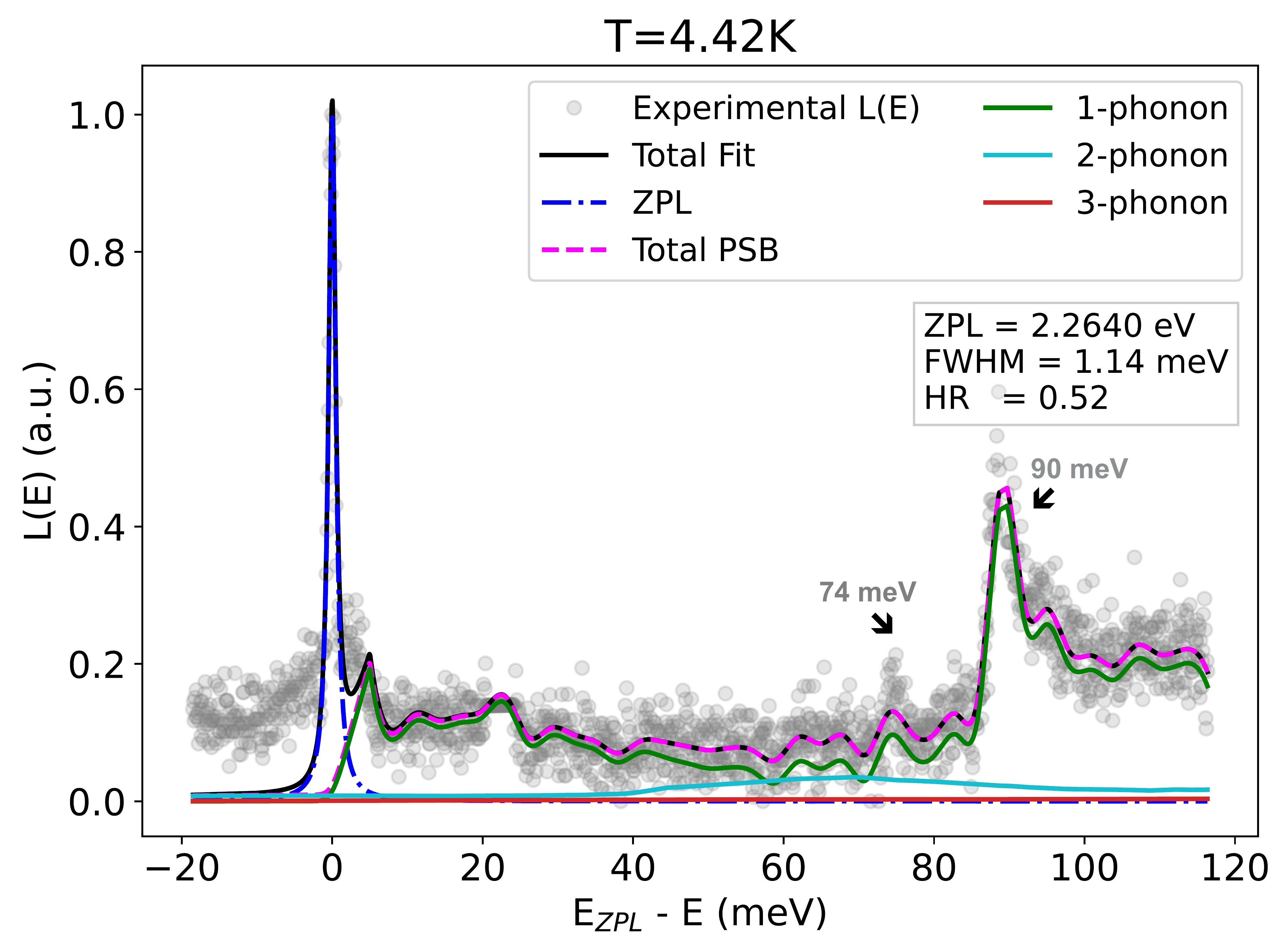}
    \caption{Temperature-dependent phonon sideband analysis at 4.42 K. Experimental spectra are shown as light-gray circles, while the overall fit is overlaid as a solid black line. The ZPL and total PSB are indicated by dashed blue and magenta curves, respectively. Individual one-, two-, and three-phonon contributions are plotted in green, cyan, and red.}
    \label{PSB_temp}
\end{figure*}

\end{appendix}